# Stability Diagram of Layer-polarized Quantum Hall States in Twisted Trilayer Graphene

Konstantin Davydov[1], Daochen Long[1], Jack A. Tavakley[1], Kenji Watanabe[2], Takashi Taniguchi[3], Ke Wang[1*]

[1]*School of Physics and Astronomy, University of Minnesota, Minneapolis, Minnesota 55455, USA*
[2]*Research Center for Electronic and Optical Materials, National Institute for Materials Science, 1-1 Namiki, Tsukuba 305-0044, Japan*
[3]*Research Center for Materials Nanoarchitectonics, National Institute for Materials Science, 1-1 Namiki, Tsukuba 305-0044, Japan*

**In the twisted trilayer graphene (tTLG) platform, the rich beating patterns between the three graphene layers give rise to a plethora of new length scales and reconstructed electronic bands arising from the emergent moiré and moiré-of-moiré superlattices. The co-existing lattices and superlattices interact and compete with each other to determine the overall transport properties of tTLG, the hierarchy of which can be electrostatically controlled by tuning the out-of-plane charge distribution or layer polarization. In this work, we measure the stability diagram of layer-polarized quantum Hall states in tTLG by systematically mapping out layer-specific Chern numbers in each layer, and intra- and interlayer Chern transitions as a function of displacement field $D$ and total carrier density $n$. In contrast to twisted bilayer systems, the rich interplay between the three atomic layers gives rise to a complex layer-polarized stability diagram with unconventional transport features that evolve rapidly with electric and magnetic fields. The stability diagram quantitatively characterizes the interlayer screening and charge distribution in tTLG with implication of strong inter-atomic-layer Coulomb coupling. Our work provides comprehensive guidance and insights into predicting and controlling layer-polarization and interlayer transitions in tTLG, and for tuning the individual role and interactions of each participating constituent towards novel material properties.**

When two pieces of 2D materials are transferred on top of each other with a twist angle, the beating pattern of the two misaligned lattices gives rise to a new lattice periodicity known as moiré superlattice. The reconstructed electronic band can be versatilely tuned by material choices and twist angles combinations, and flat-bands promoting electron correlations have been shown to give rise to a plethora of emergent quantum phenomena such as moiré superconductivity [1,2], ferromagnetism [3], correlated insulator states [4–6] and quantum anomalous Hall effect [7–11] in various twisted bilayer systems consisting of homo- or hetero- twisted-interfaces of graphene and transition metal dichalcogenides [1–14].

The twisted trilayer graphene (tTLG) system has recently attracted great research interest as a new designer material platform [15–20]. The rich interplay between the two co-existing moiré superlattices can lead to distinct atomic landscapes and emergent quantum phenomena depending on the twist angle combination. When two twist angles are alternative and equal, the two moiré superlattices are spatially aligned [15,16] to form a single atomically-reinforced moiré superlattice that is more homogenous, promoting enhanced electron correlation and more robust superconductivity with a higher critical temperature [21]. When the two twist angles are different, the atomic reconstruction between the two moiré superlattices gives rise to a higher order superlattice with a plethora of new length scales including quasi-crystalline lattice [18] and moiré-of-moiré (MoM) superlattice, with emergent new quantum phenomena including inter-moiré Hofstadter butterfly [22], anomalous quantum Hall (QH) effect [23], correlated insulator states and signatures of superconductivity at extremely low carrier density (~ $10^{10}$ cm$^{-2}$) [17].

The hierarchy of each constituting graphene lattice and moiré superlattice can be manipulated with an out-of-plane displacement field $D$, which selectively promotes/suppresses each constituting layer in interaction and competition, and thus determines the overall transport behavior. A comprehensive map of charge distribution across each graphene layer in tTLG as a function of typical experimental parameters (such as top versus bottom gate, or $n$ versus $D$) would shed important insight in unraveling the microscopic physics mechanism of emergent tTLG quantum phenomena and provide experimental guidance in versatilely tuning and designing the tTLG material properties.

A state-of-the-art electric-only Si/SiGe qubit employs a triple quantum dot (tQD) device architecture, where lateral regions (dots) of confined electrons are tunnel-coupled to each other. A charge stability diagram of tQD [24–27] maps out the charge occupation in each dot as a function of local gate voltages, as well as identifying each intra- and inter-dot charge transitions. Such comprehensive control allows versatile initialization and manipulation of single or two electron states for charge and spin qubit. Similarly, to allow informed systematic electrostatic tuning of the tTLG platform, and to identify and control the role of each underlying microscopic constituents, we study layer-polarized QH states in a twisted trilayer graphene device. We investigate the QH stability diagram of tTLG by mapping out the Chern number in each layer as a function of local gate voltages, as well as identifying each intra- and interlayer Chern transitions.

Three pieces of graphene (top/middle/bottom graphene are marked by purple/green/yellow) are stacked on top of each other with consecutive twist angles (Fig. 1a) of ~2°, which is subsequently dry etched into Hall bar geometry (see Methods) with standard 1D edge contacts (Fig. 1b). Similar to a tQD, the twist angle is chosen to be around ~2° so that each layer in tTLG is tunnel-coupled instead of strongly hybridized. The measured 4-probe resistance (Fig. 1c) at $B = 0$ exhibits high resistance states only at charge neutrality $n = 0$, without band-insulator states that are otherwise expected at $n = 9 \times 10^{12}$ cm$^{-2}$ (see SI section S3 for details) when two graphene bands hybridize. Instead, the quantum Hall states [28] become layer-polarized beyond such carrier density and exhibit complex new dependence on carrier density and displacement field under high magnetic field, similar to that of tunnel-coupled tQDs. Figure 1d shows measured longitudinal resistance $R_{xx}$ as a function of the electric field from the top ($V_{TG}$) and back ($V_{BG}$) gate voltage $E_{TG} = V_{TG}/d_t$ and $E_{BG} = V_{BG}/d_{bg}$ at $B = 7$ T, where $d_t$, $d_{bg}$ are distances from tTLG to the top, back gate respectively. At high carrier density, three sets of layer-polarized Shubnikov-de Haas (SdH) oscillations peaks are clearly visible, each with a distinct slope (of $E_{TG}$ versus $E_{BG}$) that quantitatively characterizes the gate capacitive coupling ratio of the corresponding graphene layer. The sets of SdH lines with intermediate slope of $|\Delta E_{TG}/\Delta E_{BG}| = 1$ have an equal capacitive coupling of the top and bottom layers, consistent with QH states belonging to the middle graphene layer, whose coupling to the top and bottom gates are equally screened by the top and bottom graphene. The set of the SdH lines with a slanted slope (see SI section S5 for details on determining the slope) of $|\Delta E_{TG}/\Delta E_{BG}| < 1$ ($> 1$) has a capacitive coupling ratio (see SI section S5 for details) of $|C^*_{TG}/C^*_{BG}| = 16$ (=2.2), consistent with QH states belonging to the top (bottom) graphene layer that has a stronger and unscreened coupling to the top (back) gate, and a weaker and screened coupling to the back (top) gate due to the presence of the middle and bottom (top) graphene layer. From the slopes of these slanted SdH peaks, the effective electric field after screening from two consecutive adjacent graphene layers in electric field is reduced to ~33% (~43%) of its original unscreened value. The slight difference in screening strength by the bottom (top) graphene layers can be attributed to realistic variation in atomic landscape and electronic reconstruction at the bottom (top) moiré interface.

Figure 2a shows the measured 4-probe longitudinal resistance $R_{xx}$ as a function of charge carrier density $n = (C_{TG}V_{TG} + C_{BG}V_{BG})/e + n_0$ and displacement field (positive direction defined as bottom to top) $D = (-C_{TG}V_{TG} + C_{BG}V_{BG})/2 + D_0$, where $C_{TG}$, $C_{BG}$ are the top, back gate geometric capacitances; $n_0$, $D_0$ are finite offsets (see SI section S1 for details). Figure 2b shows a zoom-in measurement of $R_{xx}$ as function of $n$ and $D$, with its Chern number configuration and intra- and interlayer QH transitions marked by the stability diagram (Figure 2c) of layer-polarized QH states. For the ease of the discussion, the same carrier density span of Figures 2b, c is also converted to the overall Landau level (LL) filling factors $\nu = nh/(eB)$ on Fig. 2b top axis, in which $n$ is the corresponding carrier density, $\varepsilon_0$ is the vacuum permittivity, $h$ is Planck's constant, and $e$ is the elementary charge. In our device, the LLs in each graphene layer have a standard four-fold degeneracy, with layer-specific Chern number (number of filled N-type Landau levels in each layer) labeled (three consecutive integers in convention of $N_T|N_M|N_B$ for the top/middle/bottom layer) in each Landau gap (Fig. 2d-m), and total Chern number $N_{tot} = N_T + N_M + N_B$ is labeled on the Fig. 2c bottom axis.

To understand the stability diagram and how charge carriers are added and distributed across each individual layer as a function of $n$ and $D$, the layer specific LLs are illustrated in Figures 2d-m for typical interlayer transitions (blue shape) and triple-points (red shape) in Figs. 2b, c. For the ease of discussion, we are using a simplified picture where the Landau level broadening is neglected, so that we can introduce the core qualitative features of the stability diagram, and leave the discussion on the microscopic details of the extended states versus localized states in disorder-broadened LLs to later part of the manuscript when their consequence on transport becomes more relevant. In this simple picture, increasing $n$ will uniformly move the Landau levels in the top/middle/bottom graphene (marked by the purple/green/yellow lines in Figs. 2d-m), by uniformly increasing the carrier density in each layer. Increasing $D$ will increase (decrease) carrier density in the bottom (top) graphene layer while keeping the overall carrier density constant, and thus moving the LLs in the bottom (top) layer down (up) with respect to the Fermi energy.

We first examine a typical configuration at the center of the 3|2|3| domain, the measured $R_{xx}$ is zero due to only ballistic edge states conducting in each layer. In the simplified picture, the Fermi level lies in the middle of the $N_T = N_B = 3$ Landau gaps (referred as $\Delta_T$, $\Delta_B$) in both the top and bottom layer, and lies in the $N_M = 2$ Landau gap (defined as $\Delta_M$) of the middle layer, $\Delta_M/8$ below the 3$^{rd}$ LL and 7$\Delta_M$ /8 above the 2$^{nd}$ LL (grey square, Fig. 2d). More accurately, this corresponds to the exact integer filling of $N = 3$ in the top and bottom layer, and filling the 3$^{rd}$ LL in the middle layer by a 3/8 of its density of states (nearly half-filled). Increasing $D$ to lower the LLs in the bottom layer and raising the LLs in the top layer by an equal amount of $\Delta_T/2 = \Delta_B/2$ (while keeping the LLs in the middle layer intact) will bring the system to a local high resistance peak corresponding to an interlayer Chern transition (blue star) between the 2|2|4 and 3|2|3 Chern number configurations (hereby denoted as 2|2|4 ⇔ 3|2|3 transition), where corresponding LLs in both top and bottom layers are half-filled (Fig. 2e) and aligned with the Fermi energy, resulting in dissipative bulk conduction, and mark the horizontal (parallel to the $n$ axis) boundary between the two QH domains in the stability diagram. Similarly, decreasing $D$ to raise the LLs in the bottom layer and lower the LLs in the top layer by an equal amount of half filling will bring the system to the interlayer Chern transition of 4|2|2 ⇔ 3|2|3 at the horizontal boundary between their QH domains (Fig. 2f, blue pentagon).

Starting from the center of 3|2|3, increasing $n$ while keeping $D = 0$ will move the LLs in all three layers down in energy (with respect to $E_F$) by an equal amount. Being closer in energy to the next unfilled LL in the middle layer, the Fermi level first aligns to the 3$^{rd}$ LL in the center layer (half-filled) while still being in the $N = 3$ Landau gaps of the top and the bottom layer (Figure 2g), at the boundary between the 3|2|3 and 3|3|3 Chern number configurations. The measured $R_{xx}$ is surprisingly zero (blue hollow triangle in Fig. 2b) at this intralayer Chern transition, which may be attributed to only the middle layer becoming dissipative while the other two layers are ballistic, in contrast to two dissipative layers at the interlayer Chern number transitions. The common corner of the vertical intralayer Chern transition of 3|2|3 ⇔ 3|3|3 and the horizontal interlayer Chern transition of 2|2|4 ⇔ 3|2|3 marks the triple point (red circle, Fig 2h) where all three layers have the LLs half-filled and dissipative, leading to a local maximum of measured $R_{xx}$.

The $N_{tot} = 8$ domains are rectangular, similar to the that observed in decoupled twisted bilayers [29–39]. In contrast, the 3|3|3 domain has a triangular shape unique to tTLG. In addition to the vertical boundary from the intralayer transition, the other two boundaries correspond to interlayer Chern transitions of 3|2|4 ⇔ 3|3|3 and 4|2|3 ⇔ 3|3|3 (blue triangle and circle, Figs. 2j, k) between adjacent graphene layers. As a function of $n$ versus $D$, the carrier density of the top and bottom layer can be tuned by increasing $n$ (new charges added to all three layers of the tTLG system) and by increasing $D$ (redistributing charges from the top layer to the bottom layer), while carrier density in the center layer is independent of $D$ and can only be tuned by overall carrier density $n$. As a result of this difference, the interlayer transitions between adjacent layers are slanted, while the interlayer transitions involving top and bottom layer are still horizontal. Along the boundary of the 3|2|4 ⇔ 3|3|3 transition, as $n$ increases, the $D$ field needed to keep the LLs in both middle and bottom layers aligned with $E_F$ (half-filled) decreases, thus the negative slope. Same for the positive slope along 4|2|3 ⇔ 3|3|3 interlayer transition.

Further increasing $n$ to $N_{tot} = 10$, the domain shape will become rectangular again, whose boundary is marked by the intralayer transition of 3|2|4 ⇔ 3|3|4 and 4|2|3 ⇔ 4|3|3 (hollow blue circle and square, Fig. 2l, m) and interlayer transition of 3|3|4 ⇔ 4|3|3 between the top and bottom layers, consistent with previous observations on different types of intra and interlayer transitions.

Figure 2o shows the measured 4-probe transverse resistance $R_{xy}$ as a function of $n$ and $D$, with the same scan range as in Figure 2a. Figure 2p shows a zoom-in measurement of $R_{xy}$, with the same scan range of Figure 2b and thus also marked by Figure 2c. A 1D cut (Figure 2r, along the dashed line I in Fig. 2p) crossing multiple quantized conductance plateau confirms the expected filling factors. The $R_{xy}$ as a function of $D$ (Fig. 2s) across the three $N_{tot} = 8$ domains (along the interpolated dashed line II in Fig. 2p) shows each domain having conductance quantized at the expected value of $G = 4(3/2 + N_{tot}e^2/h) = 38e^2/h$, confirming the total Chern number of $N_{tot} = 8$. The Hall resistance at the interlayer Chern transitions located at the domain boundaries are measured to be smaller than $1/(38e^2/h)$. Similar $R_{xy}$ behavior is observed for the $N_{tot} = 10$ domains (Fig. 2u, along the interpolated dashed line IV in Fig. 2p), but with $R_{xy}$ all the way reaching zero at the 3|3|4 ⇔ 4|3|3 interlayer transition. This unusual behavior is not expected from the individual contributions from each decoupled layer to the overall $R_{xy}$. The suppression of $R_{xy}$ is observed only at the interlayer transition between the top and bottom layers, and not for interlayer transitions between adjacent layers (Fig. 2t, along the interpolated dashed line III in Fig. 2p). This suggests the underlying mechanism for this exotic behavior may be associated with a strong Coulomb coupling and drag between the top and bottom layers with partially filled LLs and $R_{xy} = 0$ implying possible emergence of exciton condensates. Future Coulomb drag experiments with individual contacts to the top and bottom layers are needed to assertively confirm the exact mechanism and investigate its dependence on $n$ and $D$.

Due to the system's mirror symmetry against the center layer, domains centered around the $D = 0$ axis have an equal top and bottom Chern number. At the center of these domains, the top and bottom layers are exactly integer-filled, corresponding to $E_F$ in the center of the LL gap. However, depending on the magnetic field $B$, filling for middle layer LLs at the domain center can be arbitrary, leading to distinct features of the resulting stability diagram.

Figure 3a shows the measured 4-probe longitudinal resistance $R_{xx}$ at $B = 5$ T, with Figure 3b showing a zoom-in scan and the corresponding stability diagram (Figure 3c, with zoom in around 5|4|5 shown in Figure 3g). At the center of the 5|3|5 domain (grey square, Figure 3d), the Fermi level lies in the middle of the $N_T = N_B = 5$ Landau gaps ($\Delta_T = \Delta_B$) in both the top and bottom layers, similar to the previous stability diagram. In contrast, the $N_M = 4$ LL is now 1/8 filled instead of 3/8 filled, corresponding to the Fermi level $3\Delta_M/8$ below the 4$^{th}$ LL and $5\Delta_M/8$ above the 3$^{rd}$ LL in the middle layer.

As a result, at the center of the 5|3|5 domain, the 4$^{th}$ LL in the center layer is only slightly closer (by 1/8 of the corresponding gap) to the Fermi level than the 6$^{th}$ LLs in top and bottom layer. Increasing carrier density $n$ at $D = 0$ will bring the $N_M = 4$ LL in the middle layer to the Fermi level (blue triangle, Figure 3h), corresponding to the 5|3|5 ⇔ 5|4|5 intra-middle-layer transition, similar to the previous stability diagram.

In contrast to the previous stability diagram where the LL in the middle layer is significantly closer to $E_F$, the smaller advantage of middle layer here can be easily overtaken by the top (bottom) layer with a negative (positive) $D$ field, that brings the next unfilled LL in the top (bottom) layer closer to $E_F$ than that in center layer, marked by a grey pentagon (star) and depicted by Figure 3e(f). Increasing the carrier density from here at negative (positive) $D$ field will therefore bring the LL in the top (bottom) layer to $E_F$ first, corresponding to the intralayer transition 5|3|5 ⇔ 6|3|5 (5|3|5 ⇔ 5|3|6) of the top (bottom) layer instead.

The size of the triangular domain 5|4|5 is proportional to the $D$ span of the 5|3|5 ⇔ 5|4|5 intra-middle-layer transition, which is now significantly reduced compared to the previous stability diagram (Fig. 2). With finite broadening of the local high $R_{xx}$ resistance peaks from the top and bottom intralayer transitions, non-zero $R_{xx}$ resistances are observed at the entire span of the

triangular 5|4|5 domain. Similar (and smaller) triangular domains can also be found along $N_{tot}$ = 12.5 and $N_{tot}$ = 15.5.

Next to these smaller triangular domains, larger trapezoid domains (such as 5|3|6 and 6|3|5|) with the measured zero $R_{xx}$ can be found, as expected from all three layers entering ballistic QH edge transport. Further increasing carrier density will trigger an intra-middle-layer transition (i.e., 5|3|6 ⇔ 5|4|6) with $R_{xx}$ = 0, similar to the previous case, together making a large combined $R_{xx}$ = 0 region.

Figure 3l (Figure 3m) shows the measured 4-probe transverse resistance $R_{xy}$ as a function of $n$ and $D$, with the same scan range as figure 3a (Figure 3b). Figures 3n-r show a few signature interpolated 1D cuts along the dashed lines I – IV labeled in Figure 3m, respectively. The dips in $R_{xy}$ previously observed at the top-bottom interlayer transitions are now observed for the entire span of the smaller triangle domains. At the boundary of the small triangular domain 5|4|5, the finite broadening of the top-middle transition 5|4|5 ⇔ 5|3|6 and 5|4|5 ⇔ 6|3|5 into the center of the 5|4|5 domain allows indirect top-bottom transition of 5|3|6 ⇔ 5|4|5 ⇔ 6|3|5 within the 5|4|5 domain, consistent with $R_{xy}$ dips possibly arising from strong Coulomb drag between partially-filled top and bottom layers and emergence of exciton condensates. For the larger trapezoid domains with $R_{xx}$ = 0, their $R_{xy}$ is consistent with the quantized conductance expected from the corresponding $N_{tot}$.

In conclusion, we report layer-polarized quantum Hall states of a consecutively twisted tTLG device, and map out the QH stability diagram of tTLG, systematically identifying layer-specific Chern numbers, triple points, intra- and interlayer Chern transitions as a function of local gate voltages, carrier density and displacement field. We show that the interplay between QH states in the tTLG QH stability diagram is significantly more diverse in mechanism and richer in transport behaviors, compared to that of the previously studied bilayer systems. The stability diagram systematically characterizes interlayer screening, layer-polarization and interlayer carrier redistribution, providing quantitative experimental guidance and reference for electrostatically tuning of the tTLG platform. We show that the shape of the stability diagram, as well as its transport signatures, can be sensitively tuned by changing relative LL alignment across the three layers. When the top and bottom layers both have partially filled LLs at the direct interlayer transition (line-shaped) or indirect interlayer transition (triangular-shaped), we report unusual transport signatures of $R_{xy}$ = 0, that can potentially be attributed to emergence of exciton condensates due to strong interlayer Coulomb coupling. This work paves a path towards future study of layer-polarized electronics states in tTLG, and how the rich interactions and reconstructions between them can lead novel correlated and exotic quantum phenomena.

**Methods**

The twisted trilayer graphene (tTLG) stack was made by utilizing the "cut and stack" method [40]. A suitable monolayer graphene (MLG) was first characterized using optical microscopy. The MLG flake was then cut with the cantilever of an atomic force microscope (AFM, from Park Systems: model XE7) into three separate pieces. To make a tTLG stack, a poly (bisphenol A carbonate) (PC) and polydimethylsiloxane (PDMS) stamp attached to a glass slide was used to pick up a top hexagonal boron nitride (hBN) flake [41]. After that, a few-layer graphite flake (working as a top gate) was picked up followed by a middle hBN to electrically and physically isolate the top gate. The three precut MLG flakes were then sequentially picked up. After each pick up of a precut MLG piece, the rotation stage with the remaining graphene was consecutively twisted by ~ 2°. Following this, a bottom hBN flake was picked up to encapsulate the tTLG. The assembled stack was released onto a SiO$_2$(285 nm)/Si substrate at 180°C. Upon cooling down to room temperature, the substrate with the stack was successively rinsed in chloroform, acetone and isopropanol to clean the remains of PC. Next, by using the AFM, bubble-free areas of tTLG were identified to ensure the absence of strain and/or defects that could potentially reduce the device quality and compromise electrical transport. Then, Ohmic edge contacts [42] to graphene were fabricated via electron-beam lithography followed by reactive-ion etching and electron-beam evaporation of metal (Cr/Pd/Au, with thicknesses of 1 nm /5 nm/>180 nm). Finally, the tTLG region was shaped into a multiterminal device (Fig. 1b) during an additional round of electron-beam lithography and subsequent reactive-ion etching.

The electrical and magneto-transport data was taken while the device was measured in a four-probe configuration under a 10 nA current bias (with an AC frequency of 17.777 Hz) inside a Bluefors LD250 cryostat at a temperature of $T = 20$ mK. In all regions of the device, contact 1 was used as a source of the driven current, and contacts 6 were used as a drain. The longitudinal, $R_{xx}$, (transverse, $R_{xy}$) resistances discussed in the main manuscript are found from the voltage measured between contact 9 and contact 8 (contact 9 and contact 3) shown in Figure 1b. To drive the current and measure voltages between different contacts, lock-in amplifiers (Stanford Research Systems: model GS200) were used. The graphite top gate and Si back gate voltages were controlled by two DC voltage sources (Yokogawa: Model GS200 and Keithley Instruments: Model 2400 respectively).

The studied device has a shape (Fig. 1b) distinct from a perfect Hall bar geometry, thus the measured resistance under applied magnetic field is always a linear combination of longitudinal (along the driven current) and Hall resistance. To account for this, the longitudinal resistance [symmetric as $R_{xx}(B)$] is extracted from the measured resistance between two given contacts $R_{\text{measured}}(B)$] according $R_{xx}(B) = [R_{\text{measured}}(B) + R_{\text{measured}}(-B)]/2$. Similarly, the antisymmetric, as a function of magnetic field, transverse Hall resistance is found $R_{xy}(B) = [R_{\text{measured}}(B) - R_{\text{measured}}(-B)]/2$, where $B \geq 0$.

We thank B. Shklovskii and H. Goldman for helpful discussions. This work was supported by NSF DMREF Award 1922165. Portions of this work were conducted in the Minnesota Nano Center, which is supported by the National Science Foundation through the National Nanotechnology Coordinated Infrastructure (NNCI) under Award Number ECCS-2025124. Portions of the hexagonal boron nitride material used in this work were provided by K.Wat. and T. T. K.Wat. and T.T. acknowledge support from the JSPS KAKENHI (Grant Numbers 20H00354, 21H05233 and 23H02052) and World Premier International Research Center Initiative (WPI), MEXT, Japan.

[*]kewang@umn.edu

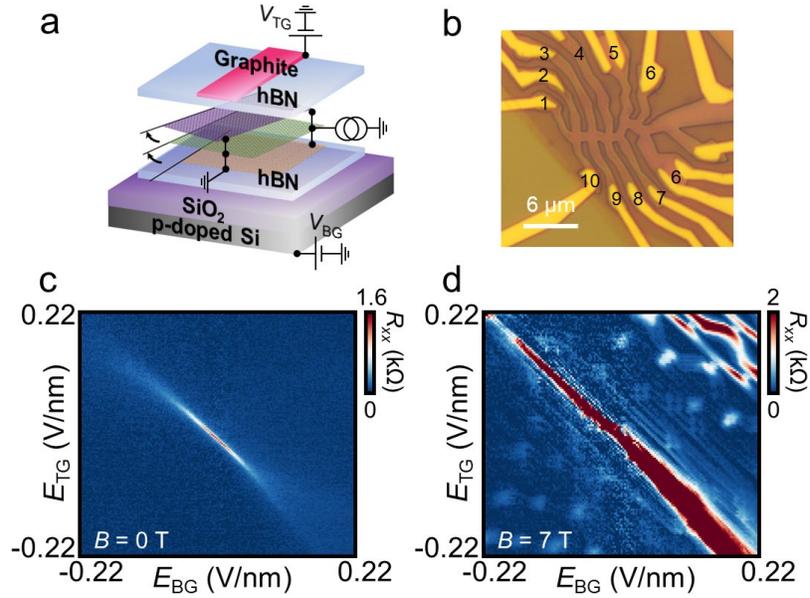

Figure 1. Gate-tunable layer coupling in tTLG. (a) Schematic architecture and measurement configuration of the-dual gated tTLG stack with each monolayer graphene (MLG) layer highlighted by a color (purple: top; green: middle; yellow: bottom). The graphite top (silicon back) gate is at applied voltage $V_{TG}$($V_{BG}$). (b) Optical image of the tTLG device used for four-probe electrical transport measurements. The metal contacts to tTLG are labeled with numbers. (c) Four-probe longitudinal resistance $R_{xx}$ as a function of the electric field from the back ($E_{BG}$) and top gate ($E_{TG}$) at $B = 0$ T. The single resistance peak corresponds to the charge neutrality point at $n = 0$ cm$^{-2}$. (d) Same as (c) but at $B = 7$ T. Additional resistance peaks are SdH maxima due to layer-polarized QH states. At $n \neq 0$ cm$^{-2}$, the resistance peaks position as a function of $E_{BG}$, $E_{TG}$ characterizes the strength of capacitive coupling between the gates and the respective layer hosting the QH states.

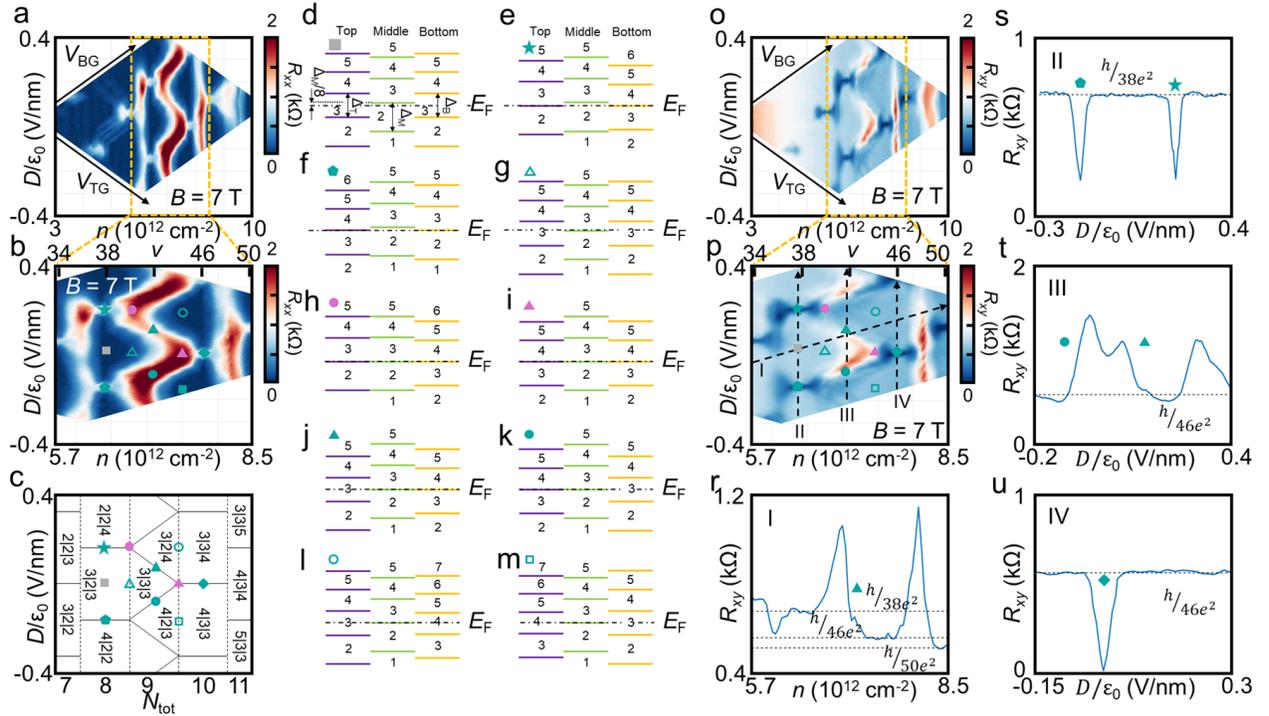

Figure 2. Stability diagram of balanced tTLG quantum Hall states. (a) Longitudinal resistance, $R_{xx}$, at changing top ($V_{TG}$) and back gate ($V_{BG}$) voltages controlling $D$ and $n$. (b) $R_{xx}$ in the highlighted $D$-$n$ area from (a). (c) Stability diagram of the layer-polarized quantum Hall states with the Chern numbers $N_T|N_M|N_B$ of the filled fourfold degenerate Landau levels in the top|middle|bottom graphene, and the total LL Chern number $N_{tot}$. The lines highlight transitions between different layer-polarized QH states involving (dashed) and without changing (solid) $N_{tot}$. (d)-(m) Energy diagrams of LLs alignment in each layer (with the color matching the layers in Fig. 1a) with respect to the Fermi energy, $E_F$, at certain QH transitions as marked in (b) and (c). Each layer-specific Landau gap is labeled with the corresponding layer Chern number. (o), (p) Hall resistance, $R_{xy}$, as a function of $D$ and $n$ in the same range as in (a), (b). (r)-(u) $R_{xy}$ linecuts from (p) labeled by the Roman numerals. $R_{xy}$ exhibits a plateau for each layer-polarized QH state with a height (highlighted by the dashed lines) consistent with the total LL filling $v$. At interlayer transitions, $R_{xy}$ behaves non-monotonically possibly due to strong Coulomb coupling and/or Coulomb drag between layers with partially filled LLs.

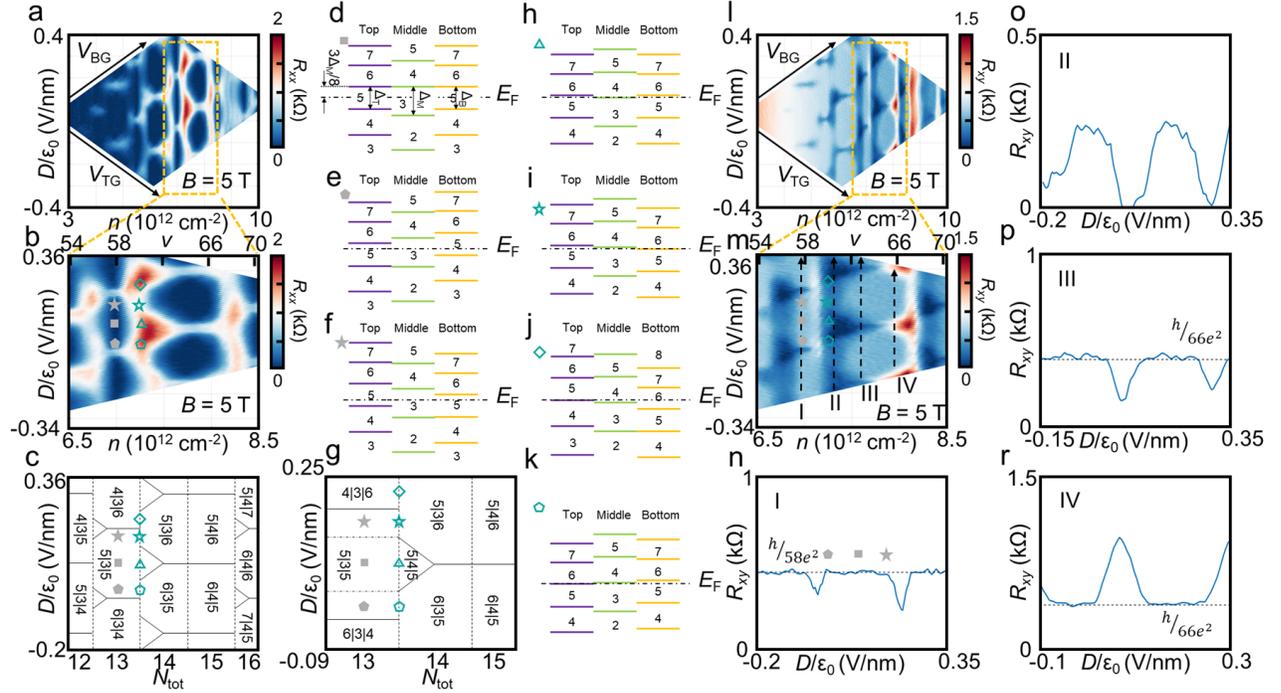

Figure 3. General stability diagram of tTLG quantum Hall states. (a) $R_{xx}$ as a function $D$ and $n$ tuned by the top ($V_{TG}$) and back gate ($V_{BG}$) voltages. (b) Zoom-in scan of the yellow area from (a) with resistance peaks corresponding to transitions between different layer-polarized QH configurations where the top axis labels the total LL filling factor $v$. (c) General stability diagram of layer-polarized quantum Hall states in the top|middle|bottom layer with the corresponding Chern numbers $N_T|N_M|N_B$ where the total Chern number is $N_{tot}$. The dashed (solid) lines trace transitions between layer-polarized QH states with (without) changing $N_{tot}$. (d)-(f) LLs alignment controlled by $D$ at a constant $n$ within the 5|3|5 layer-polarized QH state. The numbers label the layer Chern number of the Landau gap of the corresponding graphene. (g) A zoomed-in layer-polarized QH state stability diagram from (c). The markers indicate the position of the selected QH transitions on the $n$-$D$ map (b) and diagrams (c), (g). (h)-(k) LL alignment at selected transitions between the $N_{tot} = 13$ and $N_{tot} = 14$ QH states. (l), (m) $R_{xy}$ in the same range of $D$ and $n$ as in (a), (b). (n)-(r) Interpolated $R_{xy}$ linecuts [indicated in (m) and labeled by the Roman numerals] showing the $R_{xy}$ dips at interlayer transitions between layer-specific QH states.

# Stability Diagram of Layer-polarized Quantum Hall States in Twisted Trilayer Graphene

Konstantin Davydov[1], Daochen Long[1], Jack A. Tavakley[1], Kenji Watanabe[2], Takashi Taniguchi[3], Ke Wang[1*]

[1]*School of Physics and Astronomy, University of Minnesota, Minneapolis, Minnesota 55455, USA USA*
[2]*Research Center for Electronic and Optical Materials, National Institute for Materials Science, 1-1 Namiki, Tsukuba 305-0044, Japan*
[3]*Research Center for Materials Nanoarchitectonics, National Institute for Materials Science, 1-1 Namiki, Tsukuba 305-0044, Japan*

*Corresponding author. Email: kewang@umn.edu

## S1. Calculation of total charge carrier density and displacement field from gate capacitive coupling.

To calculate the total charge carrier density $n$ (that is subsequently used to find the overall Landau level filling and total Chern number) and displacement field $D$ in tTLG between the top graphite and silicon back gate, we follow a ubiquitously used (for example, in [23,43]) parallel-plate capacitor model:

$$n = \frac{C_{TG}V_{TG} + C_{BG}V_{BG}}{e} + n_0, \quad (S1)$$

$$D = \frac{-C_{TG}V_{TG} + C_{BG}V_{BG}}{2} + D_0. \quad (S2)$$

In these formulas, $V_{TG}$ ($V_{BG}$) is the voltage applied to the top (back) gate; $C_{TG}$ ($C_{BG}$) are the top(back) gate capacitances per unit area; $n_0$, $D_0$ are offset possibly due to Schottky barriers at the layer interfaces or slight intrinsic doping of graphene; $e$ is the elementary charge. The top gate geometric capacitance per unit area is calculated from $C_{TG} = \varepsilon_{hBN}/d_t$, where the hBN permittivity is $\varepsilon_{hBN} = 3.76\varepsilon_0$, $\varepsilon_0$ is the permittivity of vacuum, $d_t = 57$ nm is the thickness of hBN between the top gate and tTLG. The back gate has a capacitance of two capacitors in series with the respective separations equal the SiO$_2$ and bottom hBN (Fig. 1a) thicknesses. Thus, the geometric capacitance of the back gate is estimated according to $1/C_{BG} = 1/C_{SiO_2} + 1/C_b$, where $C_{SiO_2} = \varepsilon_{SiO_2}/d_{SiO2}$ and $C_b = \varepsilon_{hBN}/d_b$ with the SiO$_2$ permittivity $\varepsilon_{SiO_2} = 3.9\varepsilon_0$; and SiO$_2$ and bottom hBN dielectric thicknesses

$d_{SiO_2}$ = 285 nm and $d_b$ = 13 nm respectively. The total distance between the tTLG and silicon back gate introduced in the main manuscript is $d_{bg} = d_b + d_{SiO_2}$ = 298 nm. The value of $n_0 \sim 1\times10^{11}$ cm$^{-2}$ is found from the gate voltage needed to compensate for the offset of the charge neutrality resistance peak from zero charge carrier density. Same voltages give an estimate for $D_0/\varepsilon_0$ to be ~ 0.01V/nm.

## S2. Data from additional region of measured device.

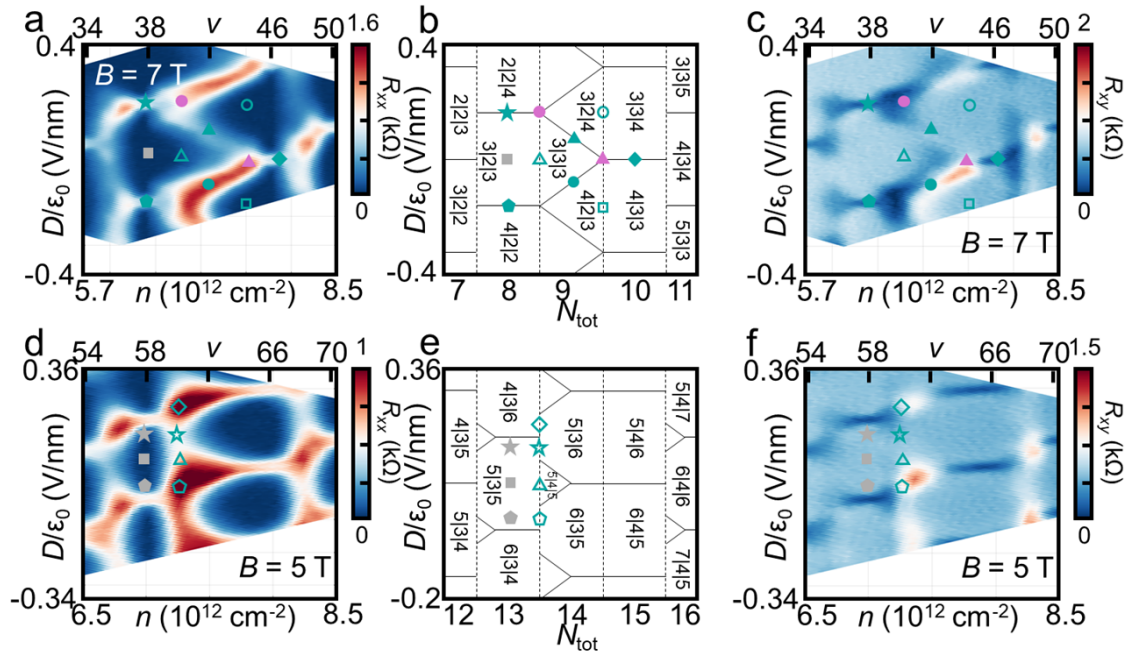

Figure S1. Results from additional region of the device. (a) Longitudinal resistance ($R_{xx}$) measured in an additional region of the device as a function of charge carrier density $n$ and displacement field $D$ at 7 T. (b) The stability diagram [outlining transitions from (a)] showing layer-polarized quantum Hall states with an integer number (matching the layer Chern number) of filled fourfold degenerate LLs in each graphene layer (top|middle|bottom = $N_T|N_M|N_B$) with the total number of filled LL $N_{tot}$. Transitions between layer-polarized QH states are traced by solid (constant $N_{tot}$) and dashed (changing $N_{tot}$) lines. Selected transitions are labeled by grey, blue and red markers of various shapes consistent with the main manuscript. (c) Transverse Hall resistance ($R_{xy}$) measured in the additional region of the device at varying $n$ and $D$ under $B$ = 7 T. (d), (e), (f) same as (a),(b),(c) but measured at $B$ = 5 T.

The longitudinal, $R_{xx}$, (transverse, $R_{xy}$) resistances presented in the main manuscript are calculated from the voltage measured between contact 9 and contact 8 (contact 9 and contact 3) under a 10 nA current driven between contact 1 and contacts 6 shown in Figure 1b. The key results discussed in the main text have been reproduced in a different region of the same device where $R_{xx}$ ($R_{xy}$) was measured between contact 8 and contact 7 (contact 8 and contact 4) under the same 10 nA current excitation between contact 1 and contacts 6.

Figure S1 summarizes magneto-transport data in the second region of the device reproducing the main features of the stability diagrams of the layer-polarized quantum Hall (QH) states described in the main manuscript. At 7 T, the measured longitudinal resistance $R_{xx}$ (Fig. S1a) as a function of $n$ and $D$ reproduces Fig. 2b with transitions between layer-polarized quantum Hall states summarized in the same stability diagram (Fig. S1b) as in the main manuscript (Fig. 2c). Each layer-polarized QH state is characterized by layer Chern numbers $N_T|N_M|N_B$ equal to the number of filled fourfold-degenerate Landau levels (LLs) in the top|middle|bottom graphene. As in the main manuscript, two types of transitions between layer-polarized QH states are observed. The first type is interlayer transitions (labeled with solid blue and red markers in Figs. S1a, b) when the Chern numbers of two selected layers are being redistributed under changing $D$ field while the total Chern number $N_{tot}$ remains the same. At these transitions, LLs of two layers are partially filled resulting in dissipative bulk conduction of two layers manifested as resistance peaks in Fig. S1a. The second type of transitions are intralayer ones (hollow markers in Figs. S1a, b) when only the middle layer becomes dissipative while its layer Chern number is changing. Figure S1a reproduces the triangular low-resistance domains corresponding to the 3|3|3, 3|2|4, 4|2|3 layer-polarized QH states being a signature feature of the tTLG system.

The behavior of $R_{xy}$ as a function of $n$ and $D$ at $B = 7$ T is also reproduced in the second region of the device (Fig. S1c). At layer-polarized QH states, $R_{xy}$ reaches a constant value $1/G_{xy}$, where $G_{xy} = 4(3/2 + N_{tot}e^2/h)$ is consistent with the layer Chern number assignment in Fig. S1b. As in the main manuscript, $R_{xy}$ changes non-monotonically at interlayer transitions reaching a local minimum at transitions involving the top and bottom layers and a maximum at transitions with two adjacent layers. As discussed in the main text, such unusual behavior signifies a non-trivial Coulomb coupling between partially filled LLs in two layers with a potential of realization of exciton condensates and Coulomb drag in the tTLG platform.

The magneto-transport data at 5 T from the main text (Fig. 3) is also mainly reproduced in the second region of the device (Figs. S1d to S1f) exhibiting the same transitions between layer-polarized QH states. However, in the second region, the triangular domains of $R_{xy}$ approaching zero are not well-developed implying a sensitive dependence of partially filled LL coupling on the exact angle combination tuning the interlayer interactions.

### S3. Characterization of two moiré periodicities.

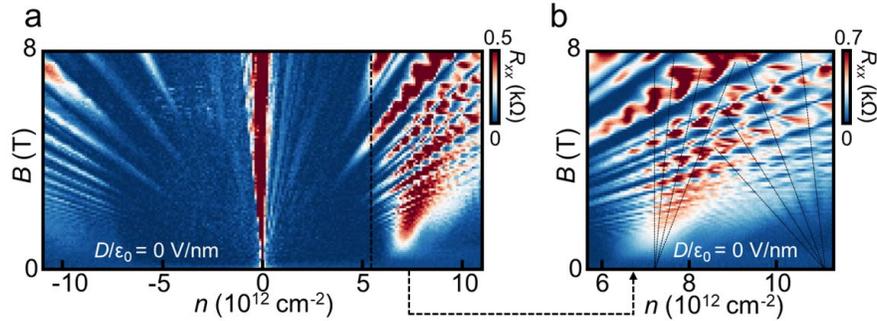

Figure S2. Effect of moiré periodicities on magneto-transport. (a) Longitudinal resistance $R_{xx}$ as a function of charge carrier density $n$ and magnetic field $B$ in the second region of the device at $D/\varepsilon_0 = 0$ V/nm. (b) Zoomed-in scan of (a) with two satellite Landau fans (highlighted by the dashed lines) due to the two moiré periodicities. The fans emanate from the charge carrier densities corresponding to four electrons per respective moiré superlattice supercell.

To confirm the three graphene layers being tunnel-coupled instead of strongly hybridized, we perform magneto-transport measurements to see the effect of the two moiré periodicities between adjacent layers. Figure S2 presents $R_{xx}$ (as an example, measured in the same region of the tTLG device as in SI section S2) as a function of magnetic field $B$ and varying charge carrier density $n$ at constant displacement field $D/\varepsilon_0 = 0$ V/nm. At $B = 0$ T, there is no signature of resistance peaks corresponding to band insulator states expected from either of the moirés to be near $n = 9\times10^{12}$ cm$^{-2}$ for the ~2° twist angles thus confirming the weak hybridization between the layers. Under finite magnetic field, in addition to the Landau fan at the charge neutrality point ($n = 0$), $R_{xx}$ exhibits a pair of satellite fans (highlighted with dashed lines in Fig. S2b, a zoom-in scan of Fig. S2a) emanating from $n_1 = 7\times10^{12}$ cm$^{-2}$ and $n_2 = 11\times10^{12}$ cm$^{-2}$ corresponding to four electrons per supercell of the two moiré interfaces. Interestingly, at hole doping, no satellite Landau fans are present. This can potentially be due to the electron-hole asymmetry of the bands of each constituent MLG layer resulting in reduced interlayer coupling and thus ill-defined moiré band gap. From the satellite fan positions, the twist angles $\theta_1 = 1.7°$ ($\theta_2 = 2.1°$) between two adjacent graphene layers can be extracted according $n_1 = 4/A_1$ ($n_2 = 4/A_2$), where $A_j = (\sqrt{3}/2)\lambda_j^2$ is the moiré supercell area, $\lambda_j = a/[2\sin(\theta_j/2)]$ is the moiré supercell lattice constant, and $a = 0.246$ nm is the graphene lattice constant, $j = 1, 2$ is the moiré interface index. The verified values of the twist angles being sufficiently away from the magic angle (where a strong hybridization between the layers is expected) gives additional evidence of weak tunnel coupling between the individual graphene monolayers in tTLG.

## S4. Electrostatic simulation of layer Landau level fillings.

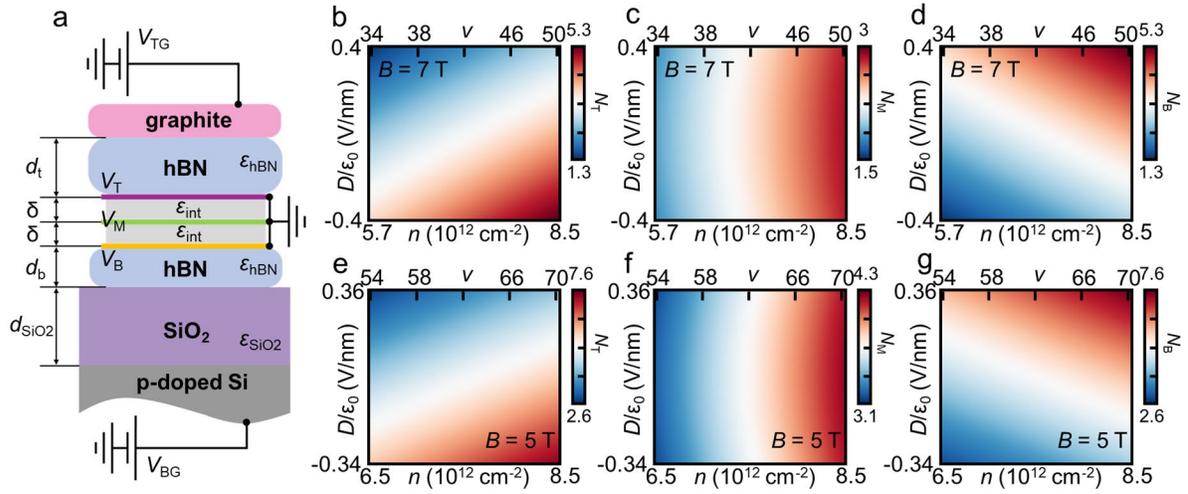

Figure S3. Electrostatic simulation of Landau level filling in individual layers of graphene. (a) Geometry of the five-plate capacitor model with the graphite top (silicon back) gate having a voltage of $V_{TG}$ ($V_{BG}$) while the three graphene layers are grounded. (b)-(d) Simulated charge carrier density in the top (b), middle (c), bottom (d) graphene converted into the corresponding layer-specific Chern numbers $N_i$ as a function of total charge carrier density $n$ and displacement field $D/\varepsilon_0$ at $B = 7$ T, where $i = $ T, M, B are the indices corresponding to the top, middle and bottom layer. (e), (f), (g) same as (b), (c), (d) but the layer Chern numbers are calculated at $B = 5$ T.

To confirm the fillings of layer-specific fourfold degenerate Landau levels discussed in the main manuscript (Figs. 2c, 3c, 3g), we perform a simulation of layer charge carrier densities by following an established modelling [43] of dual-gated twisted trilayer graphene as a five-plate capacitor (Fig. S3a). Within the model, adjacent layers of the tTLG have a vertical interlayer distance of $\delta = 0.33$ nm, the value estimated in the previous work [43] on twisted graphene. In the simulation, we neglect the second-order effect of the different top and bottom moiré twist angles on the interlayer distances in tTLG and thus assume the graphene layers being equally separated. The top graphite gate (with an electric potential $V_{TG}$) is separated from the three grounded graphene layers by an hBN layer with a thickness of $d_t = 57$ nm. The silicon back gate (at an electric potential $V_{BG}$) is isolated from the tTLG by another hBN flake with a thickness of $d_b = 13$ nm, and a layer of SiO$_2$ having a thickness of $d_{SiO_2} = 285$ nm. Doping graphene electrostatically costs charging energy, and, therefore, each layer has a non-zero electric potential $V_i$ related to the layer charge carrier density $n_i$ according $n_i = \text{sign}(V_i)\left(\frac{eV_i}{\hbar v_F}\right)^2/\pi$, where $e$ is the elementary charge, $\hbar$ is the reduced Planck's constant, $v_F = 1\times10^6$ m/s is the Fermi velocity in graphene, the index $i$ = T, M, B refers to the top, middle, bottom layer.

Layer-specific charge carrier density can be calculated from the difference of displacement fields across each graphene layer (Gauss' law) given by the following equations

$$en_T = C_{TG}(V_{TG} - V_T) + C_0(V_M - V_T),$$
$$en_M = C_0(V_T + V_B - 2V_M), \quad (S3)$$
$$en_B = C_{BG}(V_{BG} - V_B) + C_0(V_M - V_B),$$

where $C_{TG} = \varepsilon_{hBN}/d_t$ is the geometric top gate capacitance per unit area; $C_{BG}$ is the geometric back gate capacitance found from the capacitance of the bottom hBN ($C_b$) and SiO$_2$ ($C_{SiO_2}$) according $1/C_{BG} = 1/C_{SiO_2} + 1/C_b$ (see SI section S1 for additional details); $C_0 = \varepsilon_{int}/\delta$ is the interlayer capacitance per unit area with an interlayer dielectric constant $\varepsilon_{int} = 2.5\varepsilon_0$ as reported in the previous studies [43,44]. The equations are consistent with the introduced formula (S1) in SI section S1 for the total charge carrier density $n = n_T + n_M + n_B = (C_{TG}V_{TG} + C_{BG}V_{BG})/e$, where the offset $n_0 = 0$ in the simulation.

We solve the equations (S3) self-consistently for a set of top and bottom gate voltages corresponding to the same range of the total charge carrier density $n$ and displacement field $D$ as in Figs. 2b, 3b to model the layer charge carrier density. To verify the assignment of LL fillings at finite magnetic field $B$ for each layer-polarized quantum Hall (QH) state in Figs. 2c, 3c, and 3g, the layer charge carrier densities are converted into layer-specific Chern numbers $N_i$. The letter are defined as $4(N_i + 1/2) = n_i h/(eB)$ characterizing the number of filled (in this section, a non-integer Chern number indicates partial LL filling) fourfold degenerate LLs in each graphene layer at $B = 7$ T (Figs. S3b–S3d corresponding to Fig. 2 of the main manuscript) and $B = 5$ T (figs. S3e–S3g corresponding to Fig. 3) where $i$ is the layer index. In layer-polarized QH states at the absence of screening [45,46] of electric field by partially filled LLs, the layer Chern numbers found from the simulation approximately match those in Fig. 2c and Figs. 3c, 3g. Nonetheless, the simulation does not cover the effects on the charge distribution in the system from screening and coupling between partially filled Landau levels, thus the modelled data does not cover the structure of transitions between layer-polarized QH states.

**S5. Characterization of electrostatic screening by constituent graphene layers.**

To characterize the degree of electric field screening [45,46] by each graphene layer, we examine the behavior of Shubnikov-de Haas (SdH) oscillations in tTLG. Figure S4 presents a zoomed-in scan of Fig. 1d from the main manuscript at high electron doping where longitudinal resistance $R_{xx}$ at $B = 7$ T is plotted as a function of electric field due to the back ($E_{BG} = V_{BG}/d_{bg}$) and top ($E_{TG} = V_{TG}/d_t$) gate. $V_{BG}$($V_{TG}$) is the voltage applied to the back(top) gate, $d_{bg}$($d_t$) is the separation between the tTLG and the respective gate. Figure S4 shows three distinct sets (each having a characteristic slope as a function of $E_{BG}$ and $E_{TG}$) of resistance peaks (SdH oscillations maxima). The first set (traced by green dashed lines in Fig. S4) is SdH maxima changing with a slope $|\Delta E_{TG}/\Delta E_{BG}| = 1$ indicates an equal coupling to the top and back gate consistent with quantum Hall (QH) states residing in the middle graphene layer equally screened by the remaining two graphene layers. Peaks from the other two sets are indicated by purple (yellow) markers having slopes $|\Delta E_{TG}/\Delta E_{BG}| < (>)1$ corresponding to QH states in the top (bottom) graphene layer with reduced capacitive coupling to the back (top) gate due to electrostatic screening by the middle and bottom (top) graphene layer. The purple markers are assigned to data points with $R_{xx} > 2$ kΩ forming two clusters that are individually fitted with a linear function (purple dashed lines in Fig. S4) of $E_{BG}$ and $E_{TG}$. The average slope of the fitted lines is found to be $\Delta E_{TG}/\Delta E_{BG} = -0.33\pm0.03$. Similarly, the linear fit (yellow dashed line in Fig. S4) of the yellow markers ($R_{xx} > 1.2$ kΩ) has a slope of $\Delta E_{TG}/\Delta E_{BG} = -2.35\pm0.15$. From this, we can quantitatively estimate the effect of electric field screening by the middle and bottom (top) graphene layer reducing the electric field at the top (bottom) graphene to $|\Delta E_{TG}/\Delta E_{BG}| = 33\% \pm 3\%$ ($|\Delta E_{BG}/\Delta E_{TG}| = 43\% \pm 3\%$). The disparity in screening at the top (bottom) graphene can potentially be explained by structural and electronic microscopic differences of the bottom (top) moiré interfaces having distinct stacking order landscapes due to unequal twist angles between the layers.

The position of the SdH peaks in Fig. S4 from QH states of either the top or bottom layer is determined by a constant charge carrier density (constant Landau level filling) in the corresponding layer given by

$$C^*_{BG} V_{BG} + C^*_{TG} V_{TG} = \text{const}, \qquad (S4)$$

where $C^*_{BG}$ ($C^*_{TG}$) is the effective capacitance of the back (top) gate per unit area accounting for the screening effect. For the QH states in the bottom (top) layer $C^*_{BG} = C_{BG}$ ($C^*_{TG} = C_{TG}$), where $C_{BG}$ and $C_{TG}$ are the geometric gate capacitances defined in SI section S1. Equation (S4) for the peak positions can be rewritten in terms of electric field from the back (top) gate $E_{BG}$($E_{TG}$) and distances to the respective gates:

$$C^*_{BG} d_{bg} E_{BG} + C^*_{TG} d_t E_{TG} = \text{const}. \qquad (S5)$$

Using this equation, the ratio between the effective capacitances of the top and back gate can be estimated from the slope of the SdH maximum lines: $C^*_{BG}/C^*_{TG} = |(d_t \Delta E_{TG})/(d_{bg} \Delta E_{BG})| =$
$= (d_t/d_{bg})| \Delta E_{TG}/\Delta E_{BG}|$. By substituting the extracted slope for the QH states in the top (bottom) graphene layer, the capacitance ratio is $C^*_{BG}/C^*_{TG}$ is found to be $0.063 \pm 0.005$ ($0.45 \pm 0.03$).

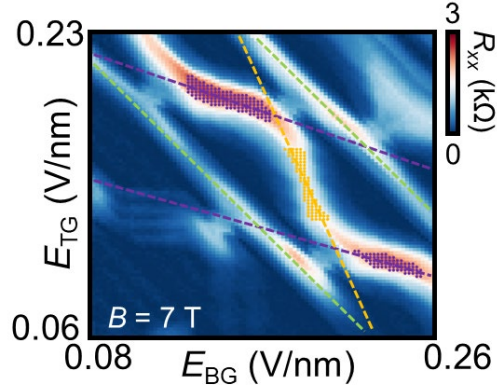

Figure S4. Characterization of electrostatic screening by graphene layers. Longitudinal resistance $R_{xx}$ as a function of electric field from the top ($E_{TG}$) and back ($E_{BG}$) gate at $B = 7$ T. The resistance peaks are Shubnikov-de Haas (SdH) oscillations maxima corresponding to quantum Hall (QH) states in different graphene layers. The peaks with a slope $|\Delta E_{TG}/\Delta E_{BG}| = 1$ (traced by the green dashed lines) correspond to QH states in the middle graphene. Peaks with slopes $|\Delta E_{TG}/\Delta E_{BG}| < (>)1$ are from QH states in the top (bottom) graphene layer with selected values of $R_{xx} > 2$ kΩ ($R_{xx} > 1.2$ kΩ) highlighted by purple (yellow) markers whose dependence on $E_{BG}$ and $E_{TG}$ is fitted with purple (yellow) lines.

## S6. Electric field screening by partially filled Landau levels.

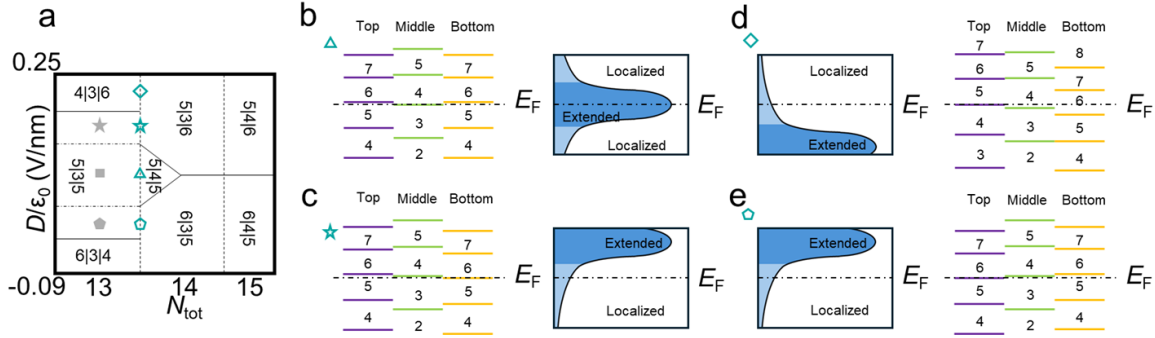

Figure S5. Effect of screening by partially filled Landau levels. (a) Stability diagram of layer-polarized QH states centered near the 5|4|5 state at $B = 5$ T. (b)-(e) Layer-specific Landau level alignment with respect to the Fermi energy $E_F$ at selected QH state transitions labeled with the hollow blue markers in (a). The cartoons on the right in (b), (c) and on the left in (d), (e) schematically show the density of states of the $N_M = 4$ middle layer Landau level and their position with respect to the Fermi energy $E_F$. The dash-dotted lines are the mobility edges separating the localized and extended QH regimes.

For a given graphene layer, charge carriers in a partially filled Landau level can significantly screen [47] external electric field while reducing its effect on the other layers. This phenomenon is particularly prominent near the 5|3|5 ⇔ 5|4|5 intra-middle-layer transition at $B = 5$ T where at zero displacement field (Fig. S5a, blue hollow triangle), the Landau levels of all three layers are closely aligned (Fig. S5b) while the Fermi level $E_F$ (Fig. S5b) is at the maximum of the middle layer 4$^{th}$ Landau level's density of states (Fig. S5b, right). At this LL alignment, even a slight displacement field can move the LLs of the top/bottom layer to the Fermi energy $E_F$ making the corresponding layer compressible thus significantly reducing the effect of electric field on the middle graphene. Specifically, while keeping the charge carrier density constant, applying a small positive displacement field (by adding negative/positive voltage to the top/back gate) leads to partial filling of the bottom layer LL (left of Fig. S5c represented by the hollow start in Fig. S5a) corresponding to the bottom graphene being in a compressible QH state. The latter screens the additional positive voltage from the back gate thus lowering the LL filling of the middle layer. As a result, the middle graphene transitions from the extended QH state at the half-filled LL (right of Fig. S5b) to a localized one (right of Fig. S5c) passing the bottom mobility edge. Further applying positive $D$ field at constant $n$, aligns the top layer LL with $E_F$ (Fig. S5d corresponding to the hollow diamond in Fig. S5a) making the top graphene compressible and screening the added negative voltage to the top gate. Thus, the absence of the additional negative top gate voltage completely fills the middle LL pushing the Fermi level through the top mobility edge (left of Fig. S5d). Similarly, a decreasing $D$ (at added positive/negative to the top/back gate) makes the top layer LL half filled (Fig. S5e, hollow pentagon in Fig. S5a), screening the increased voltage from the top gate. Consequently, the reduced positive top gate voltage moves the $E_F$ across the same bottom mobility edge as in Fig. S5c.

**REFERENCES**

[1] Y. Cao, V. Fatemi, S. Fang, K. Watanabe, T. Taniguchi, E. Kaxiras, and P. Jarillo-Herrero, Unconventional superconductivity in magic-angle graphene superlattices, Nature **556**, 43 (2018).
[2] M. Yankowitz, S. Chen, H. Polshyn, Y. Zhang, K. Watanabe, T. Taniguchi, D. Graf, A. F. Young, and C. R. Dean, Tuning superconductivity in twisted bilayer graphene, Science **363**, 1059 (2019).
[3] A. L. Sharpe, E. J. Fox, A. W. Barnard, J. Finney, K. Watanabe, T. Taniguchi, M. A. Kastner, and D. Goldhaber-Gordon, Emergent ferromagnetism near three-quarters filling in twisted bilayer graphene, Science **365**, 605 (2019).
[4] Y. Cao, V. Fatemi, A. Demir, S. Fang, S. L. Tomarken, J. Y. Luo, J. D. Sanchez-Yamagishi, K. Watanabe, T. Taniguchi, E. Kaxiras, et al., Correlated insulator behaviour at half-filling in magic-angle graphene superlattices, Nature **556**, 80 (2018).
[5] K. P. Nuckolls, M. Oh, D. Wong, B. Lian, K. Watanabe, T. Taniguchi, B. A. Bernevig, and A. Yazdani, Strongly correlated Chern insulators in magic-angle twisted bilayer graphene, Nature **588**, 610 (2020).
[6] S. Wu, Z. Zhang, K. Watanabe, T. Taniguchi, and E. Y. Andrei, Chern insulators, van Hove singularities and topological flat bands in magic-angle twisted bilayer graphene, Nat. Mater. **20**, 488 (2021).
[7] M. Serlin, C. L. Tschirhart, H. Polshyn, Y. Zhang, J. Zhu, K. Watanabe, T. Taniguchi, L. Balents, and A. F. Young, Intrinsic quantized anomalous Hall effect in a moiré heterostructure, Science **367**, 900 (2020).
[8] J. Cai, E. Anderson, C. Wang, X. Zhang, X. Liu, W. Holtzmann, Y. Zhang, F. Fan, T. Taniguchi, K. Watanabe, et al., Signatures of fractional quantum anomalous Hall states in twisted MoTe2, Nature **622**, 63 (2023).
[9] H. Park, J. Cai, E. Anderson, Y. Zhang, J. Zhu, X. Liu, C. Wang, W. Holtzmann, C. Hu, Z. Liu, et al., Observation of fractionally quantized anomalous Hall effect, Nature **622**, 74 (2023).
[10] F. Xu, Z. Sun, T. Jia, C. Liu, C. Xu, C. Li, Y. Gu, K. Watanabe, T. Taniguchi, B. Tong, et al., Observation of Integer and Fractional Quantum Anomalous Hall Effects in Twisted Bilayer MoTe 2, Phys. Rev. X **13**, 031037 (2023).
[11] A. P. Reddy, F. Alsallom, Y. Zhang, T. Devakul, and L. Fu, Fractional quantum anomalous Hall states in twisted bilayer MoTe 2 and WSe 2, Phys. Rev. B **108**, 085117 (2023).

[12] Y. Xia, Z. Han, K. Watanabe, T. Taniguchi, J. Shan, and K. F. Mak, Superconductivity in twisted bilayer WSe2, Nature **637**, 833 (2025).
[13] Y. Guo, J. Pack, J. Swann, L. Holtzman, M. Cothrine, K. Watanabe, T. Taniguchi, D. G. Mandrus, K. Barmak, J. Hone, et al., Superconductivity in 5.0° twisted bilayer WSe2, Nature **637**, 839 (2025).
[14] K. Kang, B. Shen, Y. Qiu, Y. Zeng, Z. Xia, K. Watanabe, T. Taniguchi, J. Shan, and K. F. Mak, Evidence of the fractional quantum spin Hall effect in moiré MoTe2, Nature **628**, 522 (2024).
[15] J. M. Park, Y. Cao, K. Watanabe, T. Taniguchi, and P. Jarillo-Herrero, Tunable strongly coupled superconductivity in magic-angle twisted trilayer graphene, Nature **590**, 249 (2021).
[16] Z. Hao, A. M. Zimmerman, P. Ledwith, E. Khalaf, D. H. Najafabadi, K. Watanabe, T. Taniguchi, A. Vishwanath, and P. Kim, Electric field–tunable superconductivity in alternating-twist magic-angle trilayer graphene, Science **371**, 1133 (2021).
[17] X. Zhang, K.-T. Tsai, Z. Zhu, W. Ren, Y. Luo, S. Carr, M. Luskin, E. Kaxiras, and K. Wang, Correlated Insulating States and Transport Signature of Superconductivity in Twisted Trilayer Graphene Superlattices, Phys. Rev. Lett. **127**, 166802 (2021).
[18] A. Uri, S. C. de la Barrera, M. T. Randeria, D. Rodan-Legrain, T. Devakul, P. J. D. Crowley, N. Paul, K. Watanabe, T. Taniguchi, R. Lifshitz, et al., Superconductivity and strong interactions in a tunable moiré quasicrystal, Nature **620**, 762 (2023).
[19] J. C. Hoke, Y. Li, Y. Hu, J. May-Mann, K. Watanabe, T. Taniguchi, T. Devakul, and B. E. Feldman, *Imaging Supermoire Relaxation and Conductive Domain Walls in Helical Trilayer Graphene*, arXiv:2410.16269.
[20] Z. Zhu, S. Carr, D. Massatt, M. Luskin, and E. Kaxiras, Twisted Trilayer Graphene: A Precisely Tunable Platform for Correlated Electrons, Phys. Rev. Lett. **125**, 116404 (2020).
[21] Y. Zhang, R. Polski, C. Lewandowski, A. Thomson, Y. Peng, Y. Choi, H. Kim, K. Watanabe, T. Taniguchi, J. Alicea, et al., Promotion of superconductivity in magic-angle graphene multilayers, Science **377**, 1538 (2022).
[22] W. Ren, K. Davydov, Z. Zhu, J. Ma, K. Watanabe, T. Taniguchi, E. Kaxiras, M. Luskin, and K. Wang, Tunable inter-moiré physics in consecutively twisted trilayer graphene, Phys. Rev. B **110**, 115404 (2024).
[23] L.-Q. Xia, S. C. de la Barrera, A. Uri, A. Sharpe, Y. H. Kwan, Z. Zhu, K. Watanabe, T. Taniguchi, D. Goldhaber-Gordon, L. Fu, et al., Topological bands and correlated states in helical trilayer graphene, Nat. Phys. **21**, 239 (2025).
[24] K. Eng, T. D. Ladd, A. Smith, M. G. Borselli, A. A. Kiselev, B. H. Fong, K. S. Holabird, T. M. Hazard, B. Huang, P. W. Deelman, et al., Isotopically enhanced triple-quantum-dot qubit, Science Advances **1**, e1500214 (2015).
[25] L. Gaudreau, G. Granger, A. Kam, G. C. Aers, S. A. Studenikin, P. Zawadzki, M. Pioro-Ladrière, Z. R. Wasilewski, and A. S. Sachrajda, Coherent control of three-spin states in a triple quantum dot, Nature Phys **8**, 54 (2012).
[26] B. Kratochwil, J. V. Koski, A. J. Landig, P. Scarlino, J. C. Abadillo-Uriel, C. Reichl, S. N. Coppersmith, W. Wegscheider, M. Friesen, A. Wallraff, et al., Charge qubit in a triple quantum dot with tunable coherence, Phys. Rev. Res. **3**, 013171 (2021).
[27] J. Z. Blumoff, A. S. Pan, T. E. Keating, R. W. Andrews, D. W. Barnes, T. L. Brecht, E. T. Croke, L. E. Euliss, J. A. Fast, C. A. C. Jackson, et al., Fast and High-Fidelity State Preparation and Measurement in Triple-Quantum-Dot Spin Qubits, PRX Quantum **3**, 010352 (2022).

[28] K. v. Klitzing, G. Dorda, and M. Pepper, New Method for High-Accuracy Determination of the Fine-Structure Constant Based on Quantized Hall Resistance, Phys. Rev. Lett. **45**, 494 (1980).

[29] J. D. Sanchez-Yamagishi, T. Taychatanapat, K. Watanabe, T. Taniguchi, A. Yacoby, and P. Jarillo-Herrero, Quantum Hall Effect, Screening, and Layer-Polarized Insulating States in Twisted Bilayer Graphene, Phys. Rev. Lett. **108**, 076601 (2012).

[30] J. D. Sanchez-Yamagishi, J. Y. Luo, A. F. Young, B. M. Hunt, K. Watanabe, T. Taniguchi, R. C. Ashoori, and P. Jarillo-Herrero, Helical edge states and fractional quantum Hall effect in a graphene electron–hole bilayer, Nature Nanotech **12**, 118 (2017).

[31] Q. Li, Y. Chen, L. Wei, H. Chen, Y. Huang, Y. Zhu, W. Zhu, D. An, J. Song, Q. Gan, et al., Strongly coupled magneto-exciton condensates in large-angle twisted double bilayer graphene, Nat Commun **15**, 5065 (2024).

[32] D. Kim, B. Kang, Y.-B. Choi, K. Watanabe, T. Taniguchi, G.-H. Lee, G. Y. Cho, and Y. Kim, Robust Interlayer-Coherent Quantum Hall States in Twisted Bilayer Graphene, Nano Lett. **23**, 163 (2023).

[33] Y. Kim, P. Moon, K. Watanabe, T. Taniguchi, and J. H. Smet, Odd Integer Quantum Hall States with Interlayer Coherence in Twisted Bilayer Graphene, Nano Lett. **21**, 4249 (2021).

[34] B. Dong, K. Zhao, K. Watanabe, T. Taniguchi, J. Lu, J. Zhao, F. Wu, J. Zhang, and Z. Han, *Quantized Landau-Level Crossing Checkerboard in Large-Angle Twisted Graphene*, arXiv:2412.03004.

[35] S. Kim, D. Kim, K. Watanabe, T. Taniguchi, J. H. Smet, and Y. Kim, Orbitally Controlled Quantum Hall States in Decoupled Two-Bilayer Graphene Sheets, Advanced Science **10**, 2300574 (2023).

[36] P. Rickhaus, F. K. de Vries, J. Zhu, E. Portoles, G. Zheng, M. Masseroni, A. Kurzmann, T. Taniguchi, K. Watanabe, A. H. MacDonald, et al., Correlated electron-hole state in twisted double-bilayer graphene, Science **373**, 1257 (2021).

[37] F. K. de Vries, J. Zhu, E. Portolés, G. Zheng, M. Masseroni, A. Kurzmann, T. Taniguchi, K. Watanabe, A. H. MacDonald, K. Ensslin, et al., Combined Minivalley and Layer Control in Twisted Double Bilayer Graphene, Phys. Rev. Lett. **125**, 176801 (2020).

[38] S. Pezzini, V. Mišeikis, G. Piccinini, S. Forti, S. Pace, R. Engelke, F. Rossella, K. Watanabe, T. Taniguchi, P. Kim, et al., 30°-Twisted Bilayer Graphene Quasicrystals from Chemical Vapor Deposition, Nano Lett. **20**, 3313 (2020).

[39] Y. Yuan, L. Liu, J. Zhu, J. Dong, Y. Chu, F. Wu, L. Du, K. Watanabe, T. Taniguchi, D. Shi, et al., Interplay of Landau Quantization and Interminivalley Scatterings in a Weakly Coupled Moiré Superlattice, Nano Lett. **24**, 6722 (2024).

[40] Y. Saito, J. Ge, K. Watanabe, T. Taniguchi, and A. F. Young, Independent superconductors and correlated insulators in twisted bilayer graphene, Nat. Phys. **16**, 926 (2020).

[41] C. R. Dean, A. F. Young, I. Meric, C. Lee, L. Wang, S. Sorgenfrei, K. Watanabe, T. Taniguchi, P. Kim, K. L. Shepard, et al., Boron nitride substrates for high-quality graphene electronics, Nature Nanotech **5**, 10 (2010).

[42] L. Wang, I. Meric, P. Y. Huang, Q. Gao, Y. Gao, H. Tran, T. Taniguchi, K. Watanabe, L. M. Campos, D. A. Muller, et al., One-Dimensional Electrical Contact to a Two-Dimensional Material, Science **342**, 614 (2013).

[43] A. Uri, S. C. de la Barrera, M. T. Randeria, D. Rodan-Legrain, T. Devakul, P. J. D. Crowley, N. Paul, K. Watanabe, T. Taniguchi, R. Lifshitz, et al., Superconductivity and strong interactions in a tunable moiré quasicrystal, Nature **620**, 762 (2023).

[44] J. D. Sanchez-Yamagishi, T. Taychatanapat, K. Watanabe, T. Taniguchi, A. Yacoby, and P. Jarillo-Herrero, Quantum Hall Effect, Screening, and Layer-Polarized Insulating States in Twisted Bilayer Graphene, Phys. Rev. Lett. **108**, 076601 (2012).

[45] L. Britnell, R. V. Gorbachev, R. Jalil, B. D. Belle, F. Schedin, A. Mishchenko, T. Georgiou, M. I. Katsnelson, L. Eaves, S. V. Morozov, et al., Field-Effect Tunneling Transistor Based on Vertical Graphene Heterostructures, Science **335**, 947 (2012).

[46] H. Rokni and W. Lu, Layer-by-Layer Insight into Electrostatic Charge Distribution of Few-Layer Graphene, Sci Rep **7**, 42821 (2017).

[47] F. Yang, A. A. Zibrov, R. Bai, T. Taniguchi, K. Watanabe, M. P. Zaletel, and A. F. Young, Experimental Determination of the Energy per Particle in Partially Filled Landau Levels, Phys. Rev. Lett. **126**, 156802 (2021).